\documentclass[11pt,a4paper]{article}
\usepackage{jheppub_kim}

\usepackage{pdflscape}
\usepackage{amsmath}
\usepackage{amssymb}
\usepackage{dcolumn}
\usepackage{bm}
\usepackage{color}
\usepackage{epsfig}
\usepackage{amsfonts}
\usepackage{graphicx}
\usepackage{subfigure}
\usepackage{dcolumn}

\newcommand{\be}{\begin{equation}}
\newcommand{\ee}{\end{equation}}
\newcommand{\bea}{\begin{eqnarray}}
\newcommand{\eea}{\end{eqnarray}}

\def\be{\begin{equation}}
\def\ee{\end{equation}}
\def\bea{\begin{eqnarray}}
\def\eea{\end{eqnarray}}

\begin{document}

\title{Effects of the hyperscaling violation and dynamical exponents on the imaginary potential and entropic force of heavy quarkonium via holography}
\author[a]{M. Kioumarsipour}
\author[a]{J. Sadeghi}

\affiliation[a]{Sciences Faculty, Department of Physics, University of Mazandaran, Babolsar, Iran}

\emailAdd{m.kioumarsipour@stu.umz.ac.ir}
\emailAdd{pouriya@ipm.ir}

\abstract{\\
The imaginary potential and entropic force are two important different mechanisms to characterize the dissociation of heavy quarkonia. In this paper, we calculate these two quantities in strongly coupled theories with anisotropic Lifshitz scaling and hyperscaling violation exponent using holographic methods. We study how the results are affected by the hyperscaling violation parameter $ \theta $ and the dynamical exponent $ z $ at finite temperature and chemical potential. Also, we investigate the effect of the chemical potential on these quantities. As a result, we find that both mechanisms show the same results: the thermal width and the dissociation length decrease as the dynamical exponent and chemical potential increase or as the hyperscaling violating parameter decreases.

}

\maketitle

\section{Introduction}
One of the newest and interesting phases of matter, known as quark-gluon plasma (QGP) \cite{q,g,p}, is produced by the heavy ion collisions in RHIC and LHC. In order to investigate the properties of QGP, non-perturbative techniques such as the AdS/CFT correspondence \cite{a,d,s} are needed. This correspondence is a powerful and practical method connecting the supergravity theories in the $ d+1 $-dimensional anti-de Sitter (AdS) spacetimes to the quantum field theories on the $ d $-dimension boundary. Although, $ \mathcal{N}=4 $ super-Yang Mills (SYM) theory does not exactly correspond to QCD at zero temperature due to the supersymmetry and conformal invariance, they are similar just at large temperatures, at least above the critical temperature, since $ \mathcal{N}=4 $ is non-confining. The dual field theory with the AdS geometry is a conformal invariant theory, but the real QCD is not, so the conformal invariance of AdS$ _{5} $ must be broken somehow for approximating to better understanding the real QCD. To achieve this goal, many efforts have been done, such as hard wall \cite{h1} and soft wall \cite{s1} models. In addition, people have introduced the geometries that are not asymptotically AdS resulting non-conformal invariant dual filed theories that are still scale invariant. An important example is Lifshitz fixed point. Such system is anisotropic in the time direction and isotropic in spatial directions. Also, this system needs some dynamical component such as $ z $ in the time direction and is scaled invariantly under the following transformations,
\begin{equation}
(t,\mathbf{x})\rightarrow (\lambda^{z}t,\lambda \mathbf{x}).
\end{equation}
The theory is relativistic scale invariant if $z=1$, and if $z\neq 1$ the theory is non-relativistic, since the time and space directions are scaled differently. By adding the Abelian gauge and scalar fields to the Einstein gravity \cite{AA,AB}, the full class of metrics that shows scaling properties is made \cite{l1,l3,l4,l5}. Therefore, regarding both fields one can achieve the following relation,
\begin{equation}
ds^{2}=r^{-2\theta /d}\left( -r^{2z}dt^{2}+\dfrac{dr^{2}}{r^{2}}+r^{2}d{\bf x}^{2}\right), \label{tt}
\end{equation}
where $ \theta $ denotes the hyperscaling violation parameter. The metric is spatially homogeneous and has the following scale transformations,
\begin{equation}
t\rightarrow\lambda^{z}t,~~~~~r\rightarrow\lambda^{-1}r,~~~~~x_{i}\rightarrow\lambda x_{i},~~~~~ds_{d+2}\rightarrow\lambda^{\theta/d}ds_{d+2}.
\end{equation}
In order to obtain a metric having both a hyperscaling violation exponent $ \theta $ and a dynamical exponent $ z $, one can start from the Einstein-Maxwell-Dilaton (EMD) system. However, this method is not the only way and there are many other theories that can have a metric with the hyperscaling violation parameter and the dynamical exponent. For example, in ref \cite{1209}, the authors considered black branes in arbitrary dimensions. They generalized the results of ref \cite{1205.0412} for scalar black 2-branes to branes of arbitrary dimensions. In ref \cite{1212.3263} was introduced the Einstein-Proca-dilaton model in which gravity is coupled to a massive vector field and a scalar field. It showed a class of models with hyperscaling violating Lifshitz solutions having both dilatonic and axionic scalars. Also, a hyperscaling geometry may be obtained from a non-relativistic action such as Ho\v{r}ava-Lifshitz gravity with a small modification, assuming an r-dependent $ \Lambda $ or adding a non-relativistic matter \cite{1212.4190}. Hyperscaling violating solutions can be also made by some ways such as: Einstein-Maxwell-dilaton gravity with $ R^{2} $ corrections \cite{1312.2261}, coupled dilaton-squared curvature gravity \cite{1404.5399}, a theory obtained from non-supersymmetric truncations of $ \omega $-deformed $ \mathcal{N}=8 $ gauged supergravity \cite{1411.0010}, cubic gravity \cite{1508.05614}, Einstein-Maxwell-dilaton-axion model \cite{1606.07905}, a generalized Einstein-Maxwell-dilaton theory with an arbitrary number of scalars and vectors \cite{1608.03247}, a gravitational theory with two Maxwell fields, a dilatonic scalar and spatially dependent axions \cite{1608.04394} and a generalized Einstein-Maxwell-dilaton theory with an additional vector field supporting spherical or hyperbolic horizons \cite{1807.09770}. Various aspects of the holographic theories with a general dynamical exponent $ z $ and a hyperscaling violation exponent $ \theta $ were analyzed in ref \cite{null}. One of the interesting properties of such metrics is the breakage of conformal invariance. Hence, they can be suitable candidates to investigate QCD.  From the holographic viewpoint, these theories should be considered as a toy model, since they are intrinsically non-relativistic. In \cite{tay} is presented a useful review of the holographic modelling of field theories with Lifshitz symmetry. They studied a holographic dictionary for Lifshitz backgrounds, gravity duals for non-relativistic field theories, operator correlation functions, and the relationship between bulk fields and boundary operators.\\
Hyperscaling violating geometries with Lifshitz-like scaling that are solutions of Einstein-Maxwell-dilaton models are supported by running dilatonic scalar \cite{arXiv:1005.4690,arXiv:1107.2116,arXiv:0905.3337}. The existence of such running scalar shows that the solutions are not trusted in the deep IR, i.e. they are IR incomplete \cite{arXiv:1305.3279,arXiv:1208.1752}. Also, the Lifshitz and hyperscaling violation spacetimes with generic $ z $ and $ \theta $ generically display genuine null IR singularities with diverging tidal forces \cite{AA,o1,o2}. These singularities need stringy effects to be resolved \cite{o3,o5}. Also, the strings that propagate through the singularity become infinitely excited, implying the singularity cannot be resolved by stringy effects and may be improved to a full-fledged \textit{stringularity} \cite{o2,o3}.
Hence, these solutions in general do not provide a good expression of the geometry in the deep IR, so they should be modified. In other words, one needs the RG flow which connects two different spacetimes, AdS in the UV, and a non-relativistic theory with a hyperscaling violation exponent $ \theta $ and a dynamical exponent $ z $ below a certain energy scale. Ref \cite{1708} described RG flows from a conformal field theory in the UV to the IR with generic $ z $ and $ \theta $. Also, in \cite{o8}, the authors introduced novel RG flows as so-called \textit{boomerang RG flows} which flow from the $AdS_{4}$ solutions in the UV to the same $AdS_{4}$ solutions in the IR. The only difference between these two solutions is a renormalization of relative length scales. They showed that if values of deformations are large sufficiently, on the way to IR the RG flow solutions approach to two distinct intermediate scaling regimes exhibiting hyperscaling violation.
In fact, the Lifshitz-type geometries appear as an effective holographic explanation of some strongly coupled QFT which are valid below a given specific energy scale of the theory. Also, the hyperscaling violating theories are considered as an effective theory appearing as a low energy description of some UV complete field theory. In general, the metric (\ref{tt}) does not give a good expression in all ranges of $ r $ and must be followed by an \textit{effective} holographic approach in which the dual theory lives on a finite $ r=r_{F} $ slice, and for practical purposes one should consider the boundary theory at this scale \cite{null}. Hence, by following the effective holographic approach, these backgrounds can provide a good expression of the dual field theory only for a given range of $ r $. In the other words, the background that we consider in this paper is valid within a certain intermediate range of energies. Moreover, a partial holographic classification of UV complete quantum field theories with hyperscaling violation was viewed in ref \cite{enric}. Also, these hyperscaling violating theories are applied to study the holographic applications in the condensed matter system. These theories play an important role in the holographic description of Fermi surfaces (in this case $ r_{F} $ is associated with the Fermi momentum) \cite{arXiv:1112.0573} and logarithmic violation of the area low in the holographic entanglement entropy \cite{o3,arXiv:1112.2702}.

On the other hand, heavy quarkonium systems are proper and useful environments for examining the theoretical ideas. As known, heavy quarkonium is made by a heavy quark and a heavy antiquark. In order to study systematically such systems from QCD viewpoint, effective field theory techniques known as nonrelativistic QCD (NRQCD) are needed and useful. In fact, for heavy quarks, the velocity of the quark is $ v\ll 1 $, so that its dynamics are expressed by an effective field theory, nonrelativistic QCD. NRQCD is the effective theory related to heavy quarkonium. NRQCD has been used for decaying, spectroscopy and production of heavy quarkonia. The NRQCD formalism indicates that by starting from QCD can systematically explain heavy quarkonium spectroscopy and inclusive decays \cite{cas,sot,nor}.

Melting of quarkonia, like $J/\psi$ and excited states in the medium, is one of the main experimental signatures of forming QGP \cite{kar}. As was argued in \cite{mat}, the main mechanism responsible for this suppression is the color screening. Another reason that may be more important than screening, is the imaginary part of the potential, $ImV_{Q\overline{Q}}$, \cite{im1,im2}. The first calculation of the imaginary potential of a quarkonium by using the AdS/CFT correspondence has been done by Noronha and Dumitru for $\mathcal{N}=4$ SYM theory \cite{jon}. One can obtain the imaginary part of potential by studying the effect of thermal fluctuations that are arisen by interactions between the heavy quarks and the medium. Then, the imaginary part of potential was calculated by using this method several times. For example, the imaginary potential was studied for a static \cite{sta1,sta2} and a moving quarkonium \cite{ali}, and finite 't Hooft coupling correction \cite{fad}. For more studies see \cite{fadi,25,ins,24,28}. Also \cite{oth1} have introduced other approaches to study $ImV_{Q\overline{Q}}$ from the AdS/CFT correspondence. 

On the other hand, the entropic force is another important quantity that can be responsible for the suppression of the quarkonium in the heavy ion collisions. Before suggesting the entropic force method, it had been argued that the Debye screening effects caused the quarkonium suppression. But, the recent experimental research has shown a conflict: despite the density is larger at LHC, the charmonium suppression at LHC is weaker than at RHIC \cite{lh,lhc}. Recently, Kharzeev \cite{khar} showed the mentioned contradiction is relevant to the nature of deconfinement, and the entropic force is in charge of dissociating the quarkonium. Hashimoto et al. \cite{hash} did the first calculation of the entropic force related to the heavy quark pair by using the AdS/CFT correspondence. They showed the entropy increased with the inter-quark distance, and the peak of the entropy near the transition point related to the nature of deconfinement. After \cite{hash}, several times the entropic force was calculated by using the holographic method. For example, the entropic force was studied for a moving \cite{mov} and a rotating quarkonium \cite{rot}, and a moving dipole in a Yang-Mills like theory \cite{sara}. Also, the chemical potential effect \cite{che}, $ R^{2} $ and $ R^{4} $  corrections \cite{corr} and the effect of the deformation parameter \cite{def} on the entropic force are investigated.

As mentioned before, the conformal invariance is broken in the pure QCD, therefore, the conformal invariance of the dual gravity should be broken. One interesting method is using metrics that are scale covariant and lead to hyperscaling violating in dual field theory. Such metrics are Lifshitz like and have a dynamical critical exponent $ z $ and a hyperscaling violation parameter $ \theta $.
Hence, these Lifshitz and hyperscaling like metrics that indicate the non-relativistic theories for $ z\neq 1 $ and $ \theta\neq 0 $ can be useful to study NRQCD. These considerations motivate us to study the imaginary part of potential and the entropic force in a hyperscaling violating metric background. In this paper, we purpose to apply all the parameters of the theory and see how they affect the obtained results. A background that we consider has mass, charge, dynamical exponent $ z $ and hyperscaling violation parameter $ \theta $. We see that the imaginary potential and the entropic force can be changed by changing the values of these parameters. Furthermore, imaginary potential and entropic force are different mechanisms applied for investigating the melting of the quarkonium, so comparing them can be interesting.

This paper is organized as follows. In section \ref{sec2}, we briefly review the hyperscaling metric background. In section \ref{sec3}, we investigate the imaginary part of potential quarkonium in the hyperscaling violation background and analyze the effects of the dynamical exponent $ z $, the hyperscaling violation parameter $ \theta $ and the chemical potential $ \mu $. In the section \ref{ent}, we study the effects of these quantities on the entropic force. In the last section, we summarize our results.

\section{Background geometry}\label{sec2}
In this section, we are going to review the hyperscaling violating metric background \cite{alii}. We start with the following Einstein-Maxwell-Dilaton action,
\begin{equation}
S=-\dfrac{1}{16\pi G}\int d^{d+2}x\sqrt{-g} \left (R-\dfrac{1}{2}(\partial \phi)^{2}+V(\phi)-\dfrac{1}{4}\sum_{i=1}^{2}e^{\lambda_{i}\phi}F_{i}^{2}\right ),~~~~~V(\phi)=V_{0}e^{\gamma\phi},
\end{equation}
where $ V_{0} $, $ \lambda $ and $ \gamma $ are parameters of the model. This action admits the hyperscaling violating charged black brane solutions which is given by \cite{alii},
\begin{eqnarray}
ds^{2}&=&r^{2\alpha}\left( -r^{2z}f(r)dt^{2}+\dfrac{dr^{2}}{r^{2}f(r)}+r^{2}d{\bf x}^{2}\right), ~~~~\alpha:=-\theta /d, \label{ads}\nonumber\\
f(r)&=&1-\dfrac{m}{r^{z+d-\theta}}+\dfrac{Q^{2}}{r^{2(z+d-\theta-1)}},
\end{eqnarray}
where $ z $ and $ \theta $ denote the dynamical exponent and the hyperscaling violation parameter respectively. $ m $ and $ Q $ are related to the mass and the charge of the black brane. 

Moreover, one can obtain the following time component of the gauge field,
\begin{equation}
A_{t}=\mu\left (1-\left (\dfrac{r_{h}}{r}\right )^{z+d-\theta-2}\right ),
\end{equation}
where,
\begin{equation}
\mu=Q\sqrt{\dfrac{2(d-\theta)}{z+d-\theta-2}}e^{-\sqrt{\frac{z-1-\theta/d}{2(d-\theta)}}\phi_{0}} r_{h}^{-(z+d-\theta-2)},
\end{equation}
is the chemical potential in the dual boundary gauge theory\footnote{Note that for simplicity in the following we set $ \phi_{0}=0 $.}, and $ r_{h} $ is the radius of the horizon which can be obtained by setting $ f(r)=0 $,
\begin{equation}
r_{h}^{2(d+z-\theta -1)}-mr_{h}^{d+z-\theta -2}+Q^{2}=0.
\end{equation}
Also, one can rewrite $ f(r) $ in terms of $ r_{h} $,
\begin{equation}
f(r)=1-\dfrac{r_{h}^{z+d-\theta}}{r^{z+d-\theta}}+\dfrac{Q^{2}(r_{h}^{z+d-\theta-2}-r^{z+d-\theta-2})}{r^{2(z+d-\theta-1)}r_{h}^{z+d-\theta-2}}.
\end{equation}
In addition, the Hawking temperature is given by,
\begin{equation}
T=\dfrac{r_{h}^{z}(d+z-\theta)}{4\pi}\left( 1-\dfrac{(d+z-\theta -2)Q^{2}}{d+z-\theta}r_{h}^{2(\theta -d-z+1)}\right).\label{TT} 
\end{equation}
Hence, one can find,
\begin{equation}
\dfrac{\mu}{T}=4\pi \sqrt{\dfrac{2(d-\theta)}{z+d-\theta-2}}\left (\dfrac{Qr_{h}^{d-\theta}}{(d+z-\theta)r_{h}^{2(z+d-\theta-1)}-(d+z-\theta-2)Q^{2}}\right ).
\end{equation}

Here, we note that for obtaining a physically dual field theory, the null energy condition (NEC), $ T_{\mu\nu}N^{\mu}N^{\nu}\geq 0 $, should be satisfied \cite{null}. The following condition is obtained by applying the NEC on (\ref{ads}),
\begin{equation}
(d-\theta)(d(z-1)-\theta)\geq 0.\label{nu} 
\end{equation}

In order to have the dual gravity that is consistent, the allowed values for ($ z, \theta $) should be yielded by using this condition. In the case that $ z=1 $, theory is Lorentz invariant and (\ref{nu}) indicates $ \theta \leq 0 $ or $ \theta \geq d $. In the gravity side, there exists some instabilities in the range of $ \theta > d $. The range $ \theta>d $ is just allowed due to the NEC \cite{null}, hence, we consider the range of $ \theta\leq d $. By considering the range of $ \theta\leq d $ and the condition (\ref{nu}), we can find the allowed range of the dynamical exponent, 
\begin{equation}
z\geq 1+\frac{\theta}{d}. \label{b} 
\end{equation}
Moreover, the case $ \theta =0 $ provides a scale invariant theory and in this case $ z\geq 1 $.

To proceed we should notice an important point. As noticed, the geometries with nontrivial hyperscaling violation appear as an \textit{effective} holographic expression of some strongly coupled QFT that are valid below the given specific energy scale of the theory. So, the metric background (\ref{ads}) was not assumed all the way to the boundary and it provides a good expression of the dual field theory that lives on a background spacetime shown with a surface constant of $ r_{F} $. This means that in our analysis the \textit{boundary} QFT is placed at an effective UV cut off.

\section{Imaginary potential}\label{sec3}

In this section, applying the method introduced in Refs. \cite{sta1,sta2}, we want to investigate the imaginary potential of a quark-antiquark pair in the hyperscaling violation metric background \cite{alii}. One should consider a rectangular Wilson loop along the time $ \mathcal{T} $ and spatial extension L. A dipole is placed along the spatial direction and the quarks are situated at $ x_{1}=\pm L/2 $. The parameterized coordinate is given by,
\begin{equation}
t=\tau,~~~~ x_{1}=\sigma,~~~~ x_{2}=0,~~~~x_{3}=0,~~~~r=r(\sigma).
\label{par} 
\end{equation}

To proceed, we start with the following Nambu-Goto action,
\begin{equation}
S=-\dfrac{1}{2\pi\alpha^{\prime}}\int d\tau d\sigma \mathcal{L} =-\dfrac{1}{2\pi\alpha^{\prime}}\int d\tau d\sigma \sqrt{-g},\label{NG}
\end{equation}
where $ g $ indicates the determinant of the induced metric of the string worldsheet,
\begin{equation}
g_{\alpha\beta}=G_{\mu\nu}\partial_{\alpha}X^{\mu} \partial_{\beta}X^{\nu},
\end{equation} 
where $ G_{\mu\nu} $ and $ X^{\mu} $ are the metric and the target space coordinates respectively.

The induced metric is calculated by substituting (\ref{par}) into (\ref{ads}) as following,
\begin{eqnarray}
g_{00}&=&r^{2\alpha +2z}f(r),\nonumber\\
g_{11}&=&r^{2\alpha +2}\left( 1 +\dfrac{r'^{2}}{r^{4}f(r)} \right),~~~~~ r'=\frac{dr}{d\sigma}.
\end{eqnarray}
Then, we can obtain the corresponding lagrangian density as follows,
\begin{equation}
\mathcal{L}=\sqrt{a(r)+b(r)r'^{2}},\label{lag} 
\end{equation}
where,
\begin{eqnarray}
b(r)&=&r^{4\alpha+2z-2},\label{b(r)}\\
a(r)&=&r^{4\alpha+2z+2}f(r).\label{a(r)}
\end{eqnarray} 
It is clear that the lagrangian (\ref{lag}) has no $ \sigma $-dependence explicitly, so the corresponding Hamiltonian $ H $ is a constant of motion,
\begin{equation}
H=\mathcal{L}-\dfrac{\partial \mathcal{L}}{\partial r^{\prime }}r^{\prime}=constant.
\end{equation}
From the boundary condition one can find that the string configuration has a minimum at $ \sigma=0 $, i.e. $ r(0)=r_{c} $, so that $ r^{\prime}_{c}=0 $. Using this condition we obtain,
\begin{equation}
r'=\sqrt{\dfrac{a^{2}(r)-a(r)a(r_{c})}{b(r)a(r_{c})}},\label{r'}
\end{equation}
where,
\begin{eqnarray}
a(r_{c})&=&r_{c}^{4\alpha+2z+2}f(r_{c}) , \label{ac}\\
f(r_{c})&=&1-\dfrac{r_{h}^{z+d-\theta}}{r_{c}^{z+d-\theta}}+\dfrac{Q^{2}(r_{h}^{z+d-\theta-2}-r_{c}^{z+d-\theta-2})}{r_{c}^{2(z+d-\theta-1)}r_{h}^{z+d-\theta-2}},~~~~~~f_{c}:=f(r_{c}).
\end{eqnarray}
The inter-distance $ L $ of the quark and antiquark can be derived by integrating from (\ref{r'}) as,
\begin{equation}
L=2\int dr \sqrt{\dfrac{b(r)a(r_{c})}{a^{2}(r)-a(r)a(r_{c})}}.\label{LL}
\end{equation}
In order that obtain the action of the quark-antiquark pair, one should substituted (\ref{r'}) into (\ref{NG}),
\begin{equation}
S=\dfrac{\mathcal{T} }{\pi\alpha^{\prime}}\int  dr \sqrt{\dfrac{a(r)b(r)}{a(r)-a(r_{c}) }}. \label{11}
\end{equation}
This action diverges due to the contributions of the free quark-antiquark. To avoid the divergency, the self-energy of two quarks should be subtracted from (\ref{11}). The self-energy is,
\begin{equation}
S_{0}=\dfrac{\mathcal{T} }{\pi\alpha^{\prime}}\int dr \sqrt{b(r_{0})},
\end{equation}
where $ b(r_{0})=b(r\rightarrow \infty) $. Therefore, one can obtain the real part of the heavy quark potential as following \cite{mj},
\begin{equation}
ReV_{Q\overline{Q}}=\dfrac{1}{\pi\alpha'}\int_{r_{c}}^{r_{F}} dr \left( \sqrt{\dfrac{a(r)b(r)}{a(r)-a(r_{c}) }}-\sqrt{b(r_{0})}\right)-\dfrac{1}{\pi\alpha'}\int_{r_{h}}^{r_{c}}  dr \sqrt{b(r_{0})}.
\end{equation}
Finally, we use the thermal worldsheet fluctuation method \cite{jon} and obtain imaginary part of the heavy quark potential which is given by,
\begin{equation}
ImV_{Q\overline{Q}}=-\dfrac{1}{2\sqrt{2}\alpha'}\sqrt{b(r_{c})}\left(\dfrac{a'(r_{c})}{2a''(r_{c})}-\dfrac{a(r_{c})}{a'(r_{c})}\right),\label{imm} 
\end{equation}
where,
\begin{eqnarray}
b(r_{c})&=&r_{c}^{4\alpha+2z-2},\\
a'(r_{c})&=&r_{c}^{\gamma} \left(f'_{c}+\dfrac{\gamma}{r_{c}}f_{c} \right),\\
a''(r_{c})&=&r_{c}^{\gamma} \left(f''_{c}+\dfrac{2\gamma}{r_{c}}f'_{c}+\dfrac{\gamma(\gamma-1)}{r_{c}^{2}}f_{c} \right),~~~~~~~~~~
\end{eqnarray}
and,
\begin{eqnarray}
f'(r_{c})&=& \dfrac{(z+d-\theta)}{r_{c}^{z+d-\theta +1}}\left (r_{h}^{z+d-\theta}+\dfrac{Q^{2}}{r_{h}^{z+d-\theta-2}}\right )
-\dfrac{2(z+d-\theta-1)Q^{2}}{r_{c}^{2(z+d-\theta) -1}},\label{ffc}\\
f''(r_{c})&=& -\dfrac{(z+d-\theta)(z+d-\theta +1)}{r_{c}^{z+d-\theta +2}}\left (r_{h}^{z+d-\theta}+\dfrac{Q^{2}}{r_{h}^{z+d-\theta-2}}\right )\nonumber\\&&+\dfrac{2(z+d-\theta -1)(2z+2d-2\theta -1)Q^{2}}{r_{c}^{2(z+d-\theta )}},~~~~~~~~~~~\label{fff}
\end{eqnarray}
where the derivatives are with respect to $ r $ and $ \gamma:= 4\alpha+2z+2 $.

Here, we consider three restrictions on the formula (\ref{imm}) as mentioned in \cite{jon,fi}. The first restriction shows that the term $ b(r_{c}) $ must be positive, which yields,
\begin{equation}
r_{c}^{4\alpha+2z-2}> 0.
\end{equation}
where clearly is always positive.

Note that the imaginary part of potential should be negative, so the second restriction becomes,
\begin{equation}
\dfrac{a'(r_{c})}{2a''(r_{c})}-\dfrac{a(r_{c})}{a'(r_{c})} >0.
\end{equation}

\begin{figure}
\begin{center}$
\begin{array}{cc}
\includegraphics[width=73 mm]{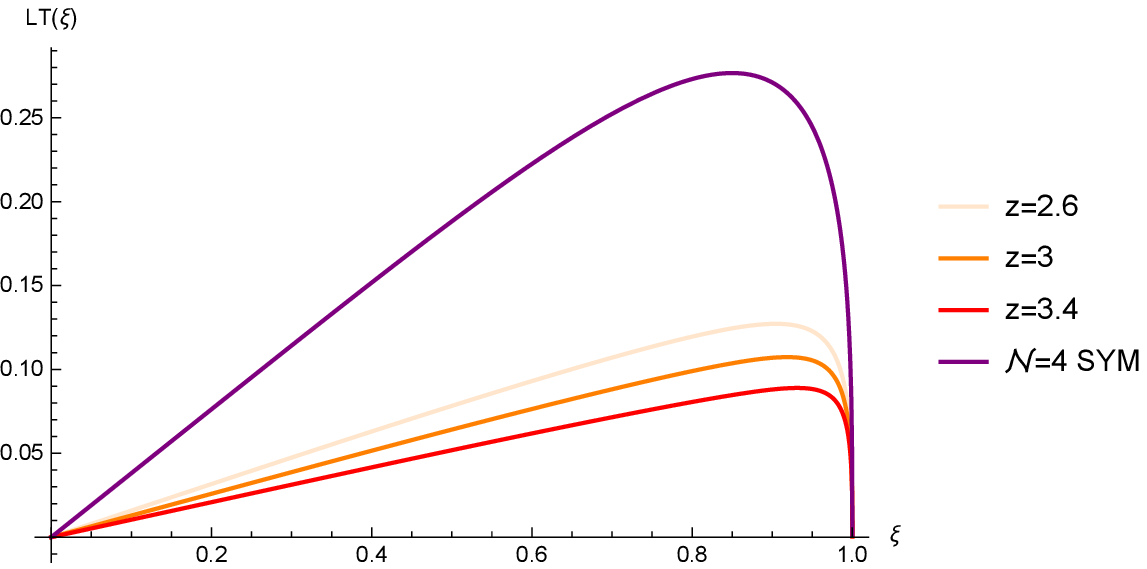}
\hspace{0.2cm}
\includegraphics[width=73 mm]{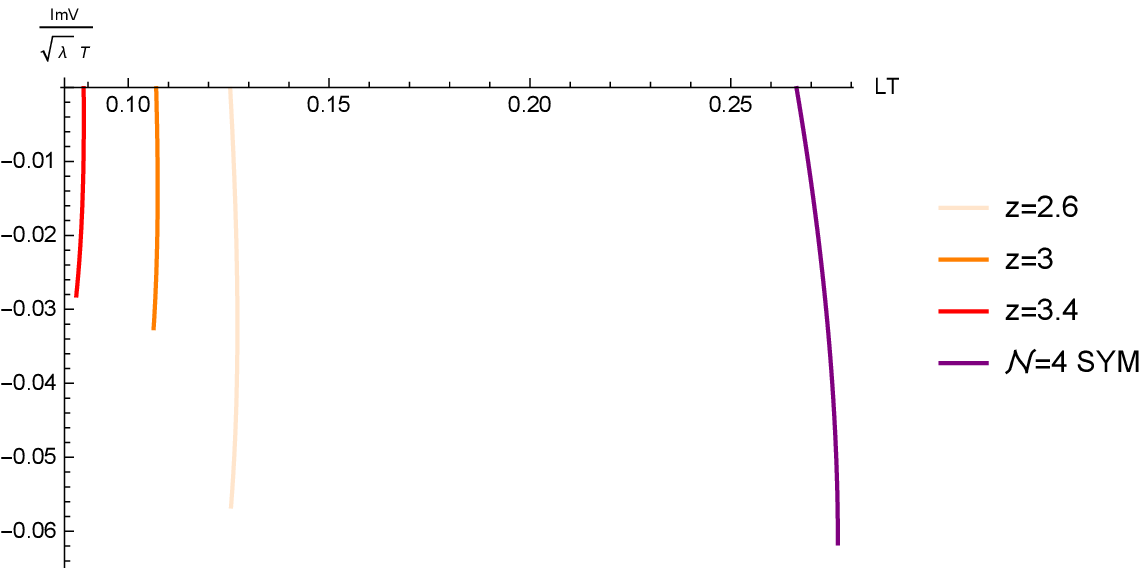}
\end{array}$
\end{center}
\caption{Left: the plot of $ LT $ versus $ \xi $ in the different $ z $ at $ \theta=2 $ and $ \mu/T=8 $. Right: $ ImV_{Q\overline{Q}}/\sqrt{\lambda}T $ versus $ LT $ in the different $ z $ at $ \theta=2 $ and $ \mu/T=8 $.}
\label{w}
\end{figure}

\begin{figure}
\begin{center}$
\begin{array}{cc}
\includegraphics[width=73 mm]{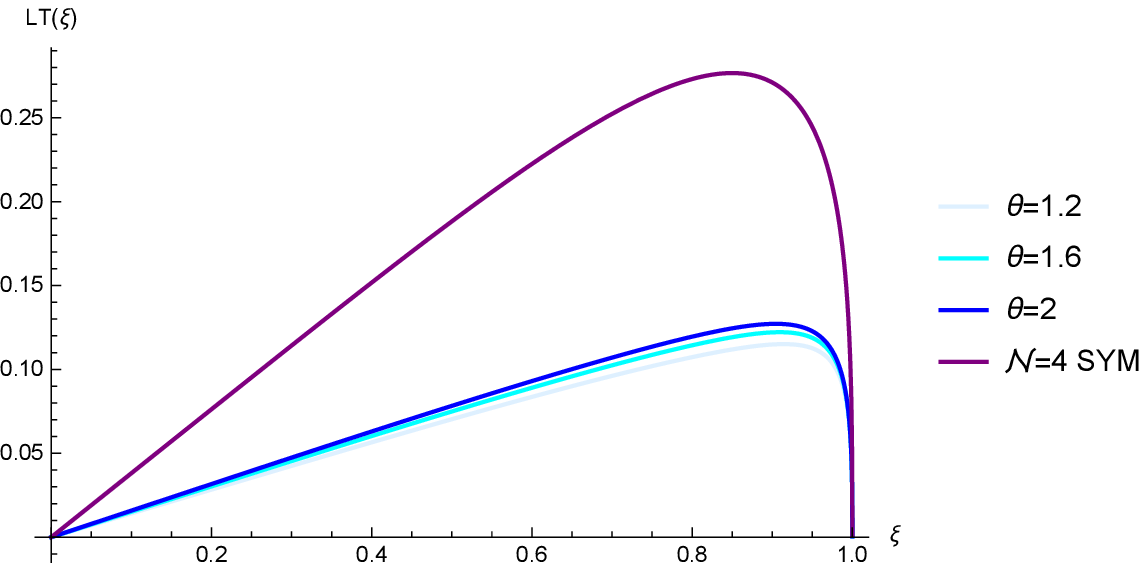}
\hspace{0.2cm}
\includegraphics[width=73 mm]{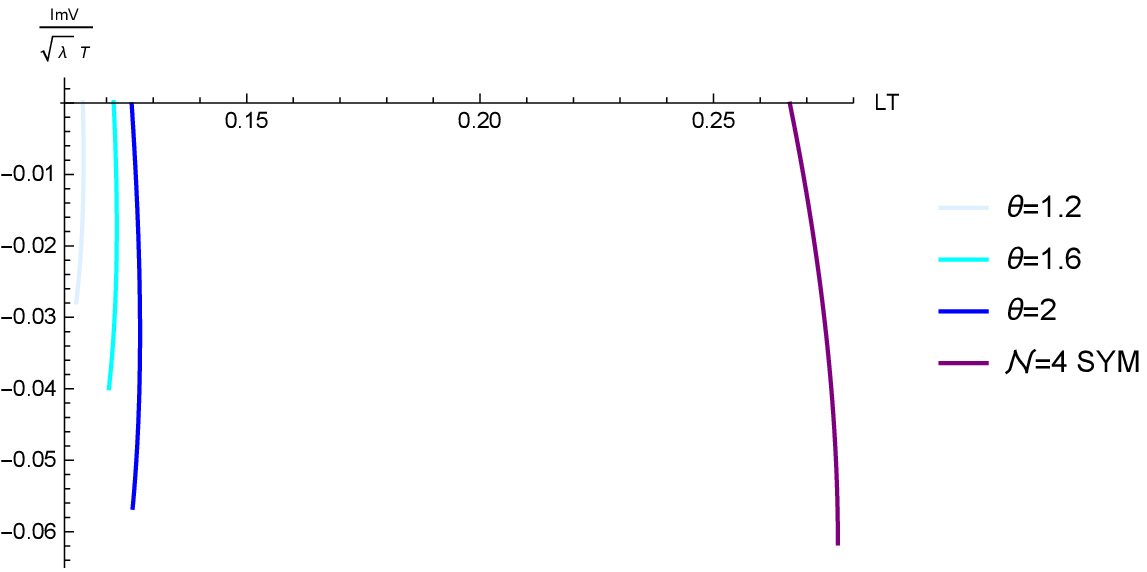}
\end{array}$
\end{center}
\caption{Left: the plot of $ LT $ versus $ \xi $ in the different $ \theta $ at $ z=2.6 $ and $ \mu/T=8 $. Right: $ ImV_{Q\overline{Q}}/\sqrt{\lambda}T $ versus $ LT $ in the different $ \theta $ at $ z=2.6 $ and $ \mu/T=8 $.}
\label{ww}
\end{figure}

\begin{figure}
\begin{center}$
\begin{array}{cc}
\includegraphics[width=73 mm]{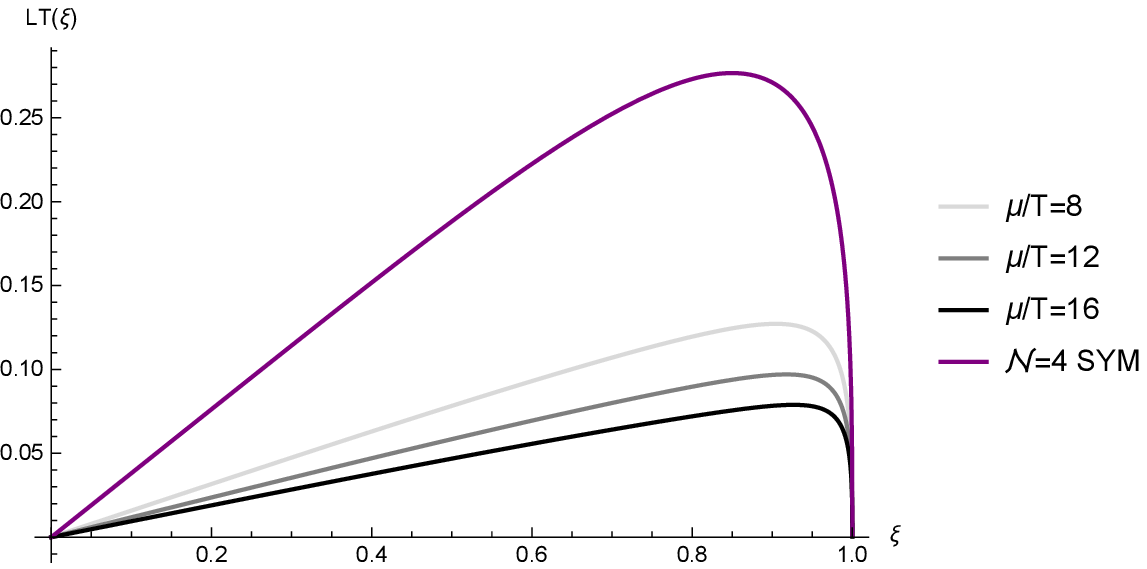}
\hspace{0.2cm}
\includegraphics[width=73 mm]{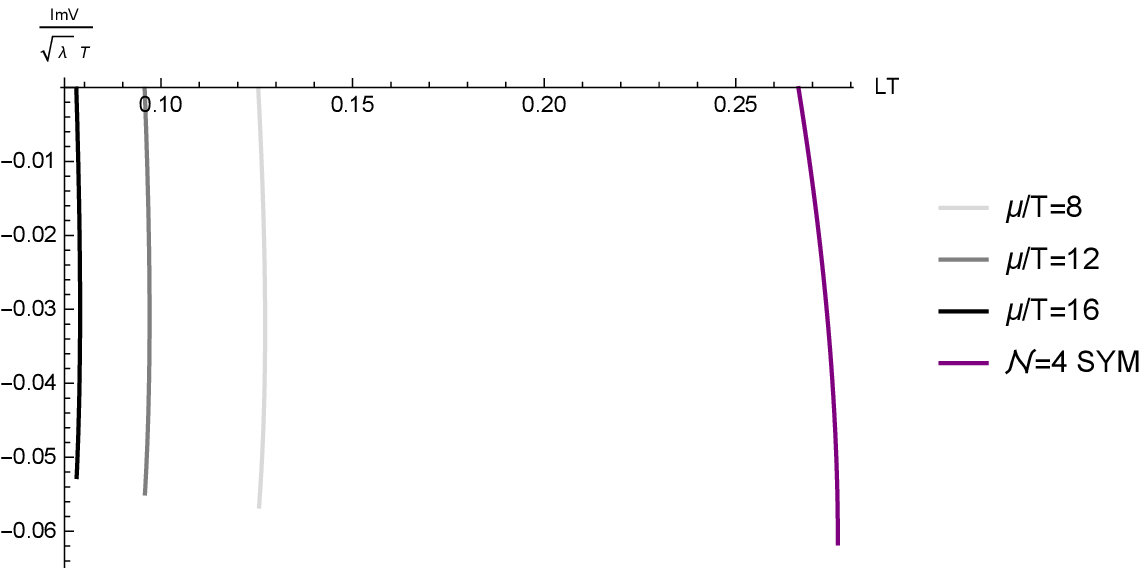}
\end{array}$
\end{center}
\caption{Left: the plot of $ LT $ versus $ \xi $ in the different $ \mu/T $ at $ \theta=2 $ and $ z=2.6 $. Right: $ ImV_{Q\overline{Q}}/\sqrt{\lambda}T $ versus $ LT $ in the different $\mu/T $ at $ \theta=2 $ and $ z=2.6 $.}
\label{www}
\end{figure}

The third restriction is relevant to the maximum value of the $ LT(\xi) $, $ LT_{max} $, which shows the limit of validity of the saddle point approximation. For this purpose, we plot numerically $ LT $ as a function of $ \xi(\equiv r_{h}/r_{c}) $ for different values $ z $, $ \theta $ and $ \mu/T $ in figures \ref{w}-\ref{www}, as well as in all plots we consider $ r_{h}=1 $. As we can see clearly, there is a maximum value of $ LT $ in each plot and it is a ascending function of $ \xi $ up to the critical value $ \xi_{max} $ and a descending function for $ \xi >\xi_{max} $. In fact, highly curved configurations which are not solutions of the Nambu-Goto action must be considered for the region of $ \xi >\xi_{max} $ \cite{bac}. Therefore, we interest to consider only the region of $ \xi < \xi_{max} $.

Now, we are going to investigate the effects of the dynamical exponent $ z $, the hyperscaling violation parameter $ \theta $ and the chemical potential $ \mu/T $ on the inter-distance and the imaginary potential. We can see that the dependency of the inter-distance and the imaginary potential to $ r $ is different and complicated in the different values $ z $ and $ \theta $, and we can not solve them analytically, therefore, we have to use numerical methods. In figure \ref{w}, we plot numerically the $ LT $ as a function of $ \xi $ and $ ImV_{Q\overline{Q}}/(\sqrt{\lambda}T) $ versus $ LT $ in the different values of $ z $. In the left panel, one can see clearly that increasing $ z $ leads to decreasing $ LT_{max} $. From the right panel, one can see that the imaginary potential begins at a $ L_{min} $ which is solution of $ ImV_{Q\overline{Q}}=0 $, and ends at a $ L_{max} $. In fact, the absolute value of the imaginary potential decreases by increasing the dynamical exponent $ z $. The dynamical exponent generates the $ ImV_{Q\overline{Q}} $ for smaller distances. Note that the suppression will be stronger if the onset of the $ ImV_{Q\overline{Q}}$ happens for smaller $LT$ \cite{fi}. Therefore, the presence of the dynamical exponent tends to decrease the thermal width and makes the suppression stronger. Interesting, the effect of the dynamical exponent $ z $ on the real potential was investigated in ref. \cite{mj}. They found increasing $ z $ leads to increasing the heavy quark potential that imply dissociation of heavy quarkonium is easy at large dynamical exponent. This result is consistent with our conclusion of the thermal width.

In figure \ref{ww}, we plot numerically the $ LT $ and $ ImV_{Q\overline{Q}}/(\sqrt{\lambda}T) $ versus $ \xi $ and $ LT $ respectively, with different $ \theta $ values. One can see clearly that $ LT_{max} $ increases by increasing $ \theta $. Moreover, the absolute value of the imaginary potential increases by increasing $ \theta $. This means that, the hyperscaling violation parameter causes to generate the $ ImV_{Q\overline{Q}} $ for larger distances. Therefore, the hyperscaling violation parameter $\theta$ has the effect of increasing the thermal width. Also, we conclude that the hyperscaling violation parameter makes the suppression weaker. It was shown in ref \cite{mj} that the hyperscaling parameter $ \theta $ affects on the heavy quark potential. Increasing $ \theta $ leads to decreasing the heavy quark potential and increasing the dissociation length making the quarkonium melt harder. This conclusion is consistent with our obtained result. 

In figure \ref{www}, the $ LT $ and $ ImV_{Q\overline{Q}}/(\sqrt{\lambda}T) $ have been plotted versus $ \xi $ and $ LT $ respectively, in the different values of $ \mu/T $. As one can see from the left panel, increasing $ \mu/T $ leads to decreasing $ LT_{max} $. Also, it is obvious from the right panel that increasing $ \mu/T $ generates the $ ImV_{Q\overline{Q}} $ for smaller distances. So, $ \mu/T $ tends to reduce the thermal width and makes the suppression stronger. The effect of the chemical potential on the imaginary potential was investigated in the different metric backgrounds \cite{che,chee1,chee2}. They concluded the chemical potential decreases the dissociation length and makes quarkonia dissociation easier that is consistent with our result.

Also, by substituting $ z=1 $, $ \theta =0 $ and $ \mu/T=0 $ into (\ref{imm}), one can recover the result of \cite{jon}. From the figures \ref{w}-\ref{www}, one can see obviously that the $ ImV_{Q\overline{Q}} $ is generated for larger distances for strongly coupled $ \mathcal{N}=4 $ SYM. Thus, the quarkonium will be dissociated harder in this case.

\section{Entropic force}\label{ent}

Here, we would like to study the effects of the dynamical exponent, the hyperscaling violation parameter and the chemical potential on the entropic force in the hyperscaling violating metric background (\ref{ads}). The entropic force does not explain any fundamental interaction and is an emergent force that can be related to the entropy $ S $ and tends to increase its entropy. As proposed in \cite{khar}, the entropic force is,
\begin{equation}
\mathcal{F}=T\dfrac{\partial S}{\partial L},
\end{equation}
where $ T $ and $ L $ are the temperature and the inter-quark distance respectively, and the entropy can be obtained by,
\begin{equation}
S=-\dfrac{\partial F}{\partial T},\label{sf}
\end{equation}
where $ F $ is the free energy of the quark-antiquark pair corresponding to the on-shell action of the fundamental string in the dual gravity geometry. Hence, in order to evaluate the entropic force, $ T $, $ L $, $ F $ and $ S $ should be calculated. 

In the previous sections, the temperature and the inter-quark distance equations are obtained as the following,
\begin{eqnarray}
T&=&\dfrac{r_{h}^{z}(d+z-\theta)}{4\pi}\left( 1-\dfrac{(d+z-\theta -2)Q^{2}}{d+z-\theta}r_{h}^{2(\theta -d-z+1)}\right),\\
L&=&2\int dr \sqrt{\dfrac{b(r)a(r_{c})}{a^{2}(r)-a(r)a(r_{c})}},
\end{eqnarray}
where $ b(r) $, $ a(r) $ and $ a(r_{c}) $ are defined in (\ref{b(r)}), (\ref{a(r)}) and (\ref{ac}). 

It is obvious from the figures \ref{w}-\ref{www}, each plot $ LT $ has a maximum value, called $ c $. If $ LT $ is larger than this maximum value, the fundamental string breaks in two pieces and the quarks are screened. In this case, the free energy and the entropy are given by,
\begin{equation}
F^{(1)}=\dfrac{1}{\pi\alpha'}\int  dr,~~~~~~S^{(1)}=\sqrt{\lambda}\theta(L-\dfrac{c}{T}).
\end{equation}
If $ LT $ is smaller than $ c $, the fundamental string is connected. Then, one can obtain the following free energy of the quark-antiquark pair, 
\begin{equation}
F=\dfrac{1}{\pi\alpha'}\int _{r_{c}}^{r_{F}} dr \sqrt{\dfrac{a(r)b(r)}{a(r)-a(r_{c})}}.
\end{equation}
Also, from (\ref{sf}) one have,
\begin{equation}
S^{(1)}=-\dfrac{\sqrt{\lambda}}{2\pi}\int _{r_{c}}^{r_{F}} dr \dfrac{\big(a'(r)b(r)+a(r)b'(r)\big) \big( a(r)-a(r_{c})\big) -a(r)b(r)\big( a'(r)-a'(r_{c})\big) }{\sqrt{a(r)b(r)\big( a(r)-a(r_{c})\big)^{3} }},\label{SS}
\end{equation}
where the derivatives are with respect to $ r_{h} $ and $ r_{F}\gg r_{c} $.

\begin{figure}
\begin{center}$
\begin{array}{cc}
\includegraphics[width=73 mm]{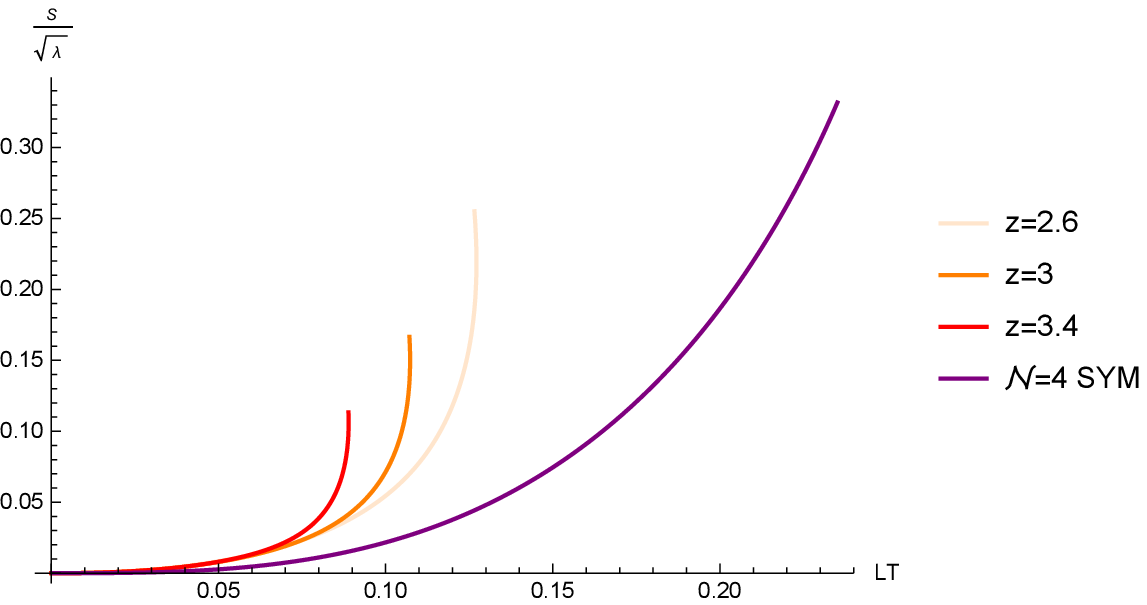}
\hspace{0.2cm}
\includegraphics[width=73 mm]{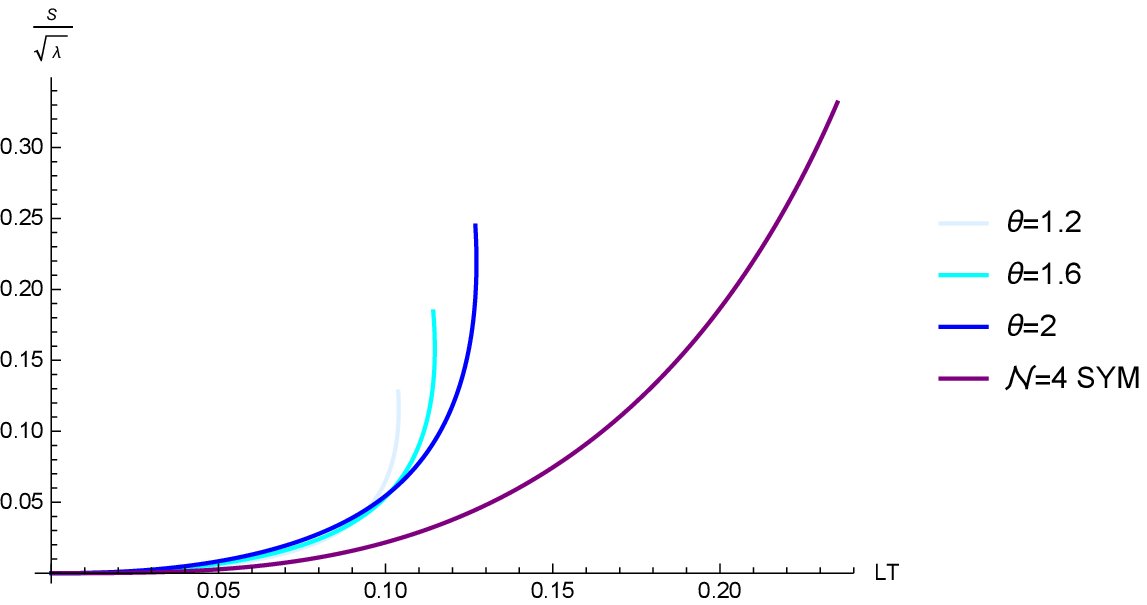}
\end{array}$
\end{center}
\caption{Left: $ S/\sqrt{\lambda} $ versus $ LT $ in different values of $ z $ at $ \theta=2 $ and $ \mu/T=8 $. Right: $ S/\sqrt{\lambda} $ versus $ LT $ in different values of $ \theta $ at $ z=2.6 $ and $ \mu/T=8 $.}
\label{entroo}
\end{figure}
\begin{figure}
\begin{center}$
\begin{array}{cc}
\includegraphics[width=73 mm]{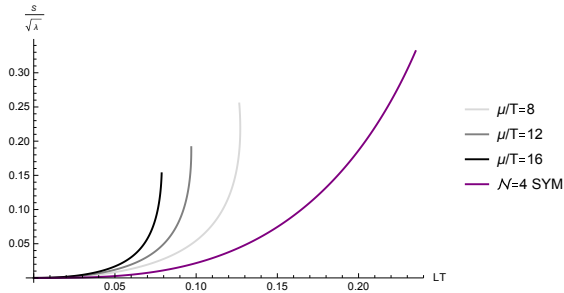}
\end{array}$
\end{center}
\caption{$ S/\sqrt{\lambda} $ versus $ LT $ in different values of $ \mu/T $ at $ \theta=2 $ and $ z=2.6 $.}
\label{ent1}
\end{figure}
In order to analyze the effects of the dynamical exponent $ z $, the hyperscaling violation parameter $ \theta $ and the chemical potential $ \mu/T $ on the entropic force, $ S/\sqrt{\lambda} $ should be plotted as a function of $ LT $ numerically. In the left panel of figure \ref{entroo}, we plot $ S/\sqrt{\lambda} $ versus $ LT $ for different $ z $ at the fixed $ \mu/T $ and $ \theta $. As we see, increasing $ z $ leads to increasing the entropy. The right plot shows the influence of the hyperscaling violation parameter on the entropy at the fixed dynamical exponent $ z $ and $ \mu/T $. It is obvious that by increasing $ \theta $ the entropy decreases. In other words, the entropy behaves differentially by increasing $ z $ and $ \theta $. In the figure \ref{ent1}, we plot $ S/\sqrt{\lambda} $ versus $ LT $ for different values of $ \mu/T $ at the fixed $ z $ and $ \theta $. We can see that the entropy increases as $ \mu/T $ increases. In ref. \cite{che}, has been investigated the effect of chemical potential on the entropic force in the AdS$ 5 $-Reissner-Nordstrom in which the entropic force increases as the chemical potential increases. Despite this metric is different ours but the chemical potential effect on the entropic force is same. As mentioned before, the entropic force that is in charge of the quarkonium destruction, is related to the growth of the entropy with the distance. So, one concludes that increasing $ z $ and $ \mu/T $ lead to increasing the entropic force. Also, the entropic force decreases by increasing $ \theta $. Then, the quarkonium will be dissociated harder by increasing $ \theta $ and easier by increasing $ z $ and $ \mu/T $. Also, in the figures \ref{entroo}-\ref{ent1} it is clear that in constant $ LT $ value, the entropic force for strongly coupled $ \mathcal{N}=4 $ SYM is less than the others.

\section{Conclusion}\label{con}

An important experimental signal for QGP formation is presented by the dissociation of a heavy quarkonium in heavy ion collisions. Two different approaches, the imaginary potential and the entropic force, have been introduced to investigate the melting the heavy quarkonium.
Because these approaches are different mechanisms for the dissociation of heavy quarkonia, thus it is interesting to compare same effects quantities on the imaginary potential with that ones on the entropic force.

In this paper, we have investigated the dynamical exponent $ z $, the hyperscaling violation parameter $ \theta $ and the chemical potential $ \mu/T $ effects on the imaginary potential and the entropic force of a quarkonium in the hyperscaling violating metric background. For imaginary potential mechanism it is shown that the dynamical exponent $ z $ and the chemical potential $ \mu/T $ tend to decrease the thermal width and make the suppression stronger; and the hyperscaling violation parameter $ \theta $ increases the thermal width and makes the suppression weaker. Thus, the dynamical exponent and the chemical potential make the quark-antiquark melt easier, while the hyperscaling violation has opposite effect. Also, we have concluded that the thermal width is larger for strongly coupled $ \mathcal{N}=4 $ SYM, thus the quarkonium dissociates harder in this case.

For the entropic force mechanism, it is shown that the dynamical exponent $ z $ and the chemical potential $ \mu/T $ increases the entropic force. The entropic force decreases by increasing the hyperscaling violation parameter $ \theta $. Thus, increasing the dynamical exponent and the chemical potential enhance the quarkonium dissociation and increasing the hyperscaling violation parameter suppresses the quarkonium dissociation. We conclude that the obtained results of these mechanism are consistent. However, it is unknown the connection between these mechanisms yet.\\
\textbf{Acknowledgements}\\
The authors are grateful very much to Dr. Bahman Khanpour for useful discussions.

\end{document}